\documentclass[journal]{IEEEtran}

\usepackage{fancyhdr}
\usepackage{amsmath}
\usepackage{amssymb}
\usepackage{graphicx}
\usepackage{booktabs}
\usepackage{amsfonts}
\usepackage{multirow}
\usepackage{algorithm}
\usepackage{algorithmic}
\usepackage{mathrsfs}
\usepackage{bm}
\usepackage{exscale}
\usepackage{relsize}
\usepackage{setspace}
\usepackage{color}
\usepackage{ulem}

\newtheorem{theorem}{\textbf{Theorem}}
\newtheorem{corollary}{\textbf{Corollary}}

\ifCLASSINFOpdf
\else
\fi

\begin{document}

\title{Towards Higher Spectral Efficiency: Spatial Path Index Modulation Improves Millimeter-Wave Hybrid Beamforming}

\author{
	Jintao~Wang,~\IEEEmembership{Senior~Member,~IEEE},
    Longzhuang~He,~\IEEEmembership{Student~Member,~IEEE},
	and Jian~Song,~\IEEEmembership{Fellow,~IEEE}

	\thanks{
        Jintao Wang, Longzhuang He and Jian Song are with the Department of Electronic Engineering, Tsinghua University, Beijing, 100084, China (e-mail: wangjintao@tsinghua.edu.cn; helz14@mails.tsinghua.edu.cn; jsong@tsinghua.edu.cn).
        
        This work was supported by the National Key R\&D Program of China (Grant No. 2017YFE0112300).
	}
}

\maketitle
\begin{abstract}
    The combination of millimeter wave (mmWave) multiple-input multiple-output (MIMO) systems and index modulation (IM) technique has recently constituted a novel form of mmWave hybrid beamforming. Lots of studies have been conducted to show that such system has the potential to outperform conventional mmWave-MIMOs with respect to spectral efficiency (SE). Most of the current works only focused on designing hybrid beamforming structures to empirically achieve higher SE performance. However, a fundamental question that whether IM technique can truly improve the SE of current mmWave hybrid beamforming is still left unanswered. Against such background, in this work we firstly extend the IM-assisted mmWave system to a more generalized version, i.e. a spatial path index modulation aided mmWave (SPIM-mmWave) system, which subsumes both IM-assisted and conventional mmWave-MIMO systems as its special cases. Based on the framework of SPIM-mmWave, a fundamental study is conducted towards a theoretic condition under which SPIM-mmWave guarantees to outperform conventional mmWave-MIMO schemes. It is demonstrated that, under a specific channel condition and noise level (which will be explicitly given by our work), SPIM-mmWave guarantees to outperform conventional mmWave-MIMO systems.
\end{abstract}
\begin{IEEEkeywords}
	Multiple-input multiple-output; millimeter wave communications; index modulation; spatial modulation; hybrid beamforming; spectral efficiency.
\end{IEEEkeywords}
\IEEEpeerreviewmaketitle

\section{Introduction}
Driven by the urgent demand for huge data traffics and ultra-dense connections in future wireless communication networks, millimeter wave (mmWave) communications \cite{rangan2014mmWave} have drawn massive research attentions during the past several years. Communication in the mmWave frequency bands has brought us many benefits. Not only is the mmWave frequency band (ranging from $30$ to $300$ GHz) much wider than the conventional micro-wave frequency band, but also the small wavelength of mmWave has also accommodated massive antennas to form the so-called massive mmWave multiple-input multiple-output (MIMO) systems \cite{rappaport2011state}\cite{han2015large}, which can drastically boost the data rate using beamforming techniques \cite{alkhateeb2014mimo}.

Meanwhile, the novel concept of spatial modulation (SM) has also been recently proposed as a new MIMO candidate technique to achieve better tradeoff between spectral efficiency (SE) and energy efficiency (EE) \cite{mesleh2008spatial}-\cite{jeganathan2008spatial}. By applying the unique information-driven random antenna-switching mechanism of SM, the MIMO transmitter can achieve higher spectral efficiency without adding more radio frequency (RF) chains. As RF components usually constitute a great amount of power consumption in a modern base station (BS), the RF-reduction benefit of SM has thus turned itself into a promising solution for future ``green'' communications. Moreover, the concept of SM belongs to an even larger family of the so-called \textit{index modulation} (IM) \cite{basar2016index}-\cite{wen2018generalized} techniques, which seeks to convey information via randomly switching the indices of activated RF components (e.g. antennas or sub-carriers), such that higher SE can be achieved with moderate number of RF chains. Therefore, in this paper, we utilize the more generalized concept of IM instead of SM to represent a unique kind of technique that encodes the information bits into the indices of RF resources. 

Driven by the low-RF-chain property of IM techniques, the combination of IM and mmWave-MIMO systems has recently drawn much research attention \cite{hemadeh201612multi}-\cite{he2018spatial}. For example, in \cite{hemadeh201612multi} and \cite{hemadeh201703reduced}, the technique of space-time shift keying (STSK, a MIMO modulation method stemming from the IM family)  was proposed in combination with mmWave orthogonal frequency division multiplexing (OFDM). In \cite{ishikawa201702TVT}, the generalized spatial modulation (GenSM) technique was combined with mmWave-MIMO systems to achieve high-SE communications in mmWave indoor line-of-sight (LoS) channels. Moreover, an IM scheme using spatial signatures was proposed for mmWave communications in \cite{luosheng201710adaptive}, in which the authors proposed to adaptively design the transmitter's analog phase-shifter network to minimize the receiver's achievable bit error rate (BER).

In \cite{he2017on} and \cite{he2018spatial}, SM techniques were further proposed in combination with mmWave-MIMO hybrid beamforming regimes. Note that hybrid beamforming (HBF) \cite{ayach2014spatially}-\cite{park2017dynamic} has been widely recognized as a key technology in future mmWave communications to handle the severe pathloss of mmWave channels. Using hybrid analog beamforming (ABF) and digital beamforming (DBF) units, mmWave-MIMO systems can achieve high-SE communications using only a few number of RF chains. In \cite{he2017on} and \cite{he2018spatial}, the authors proposed to incorporate the principle of SM into the structure of HBF. It was shown empirically that SM can further boost the SE of mmWave-MIMO HBF using the same number of RF chains.

However, the performance improvement of IM-assisted mmWave-MIMO over conventional mmWave HBF systems has only been confirmed via numerical studies, while there is no theoretic guarantee on such performance improvement. Moreover, it was also shown that in some scenarios, the SE of IM-assisted mmWave-MIMO is even outperformed by its conventional counterpart, which lacks theoretic explanations.

Against this background, in this paper we seek to provide a theoretic condition under which the application of IM guarantees to improve the SE of conventional mmWave MIMO systems. The major contributions of this paper are summarized as follows.

\begin{enumerate}
    \item Inspired by the previous IM-assisted mmWave-MIMO systems, in this paper we extend it to a more generalized version, i.e. the spatial path index modulation aided mmWave (SPIM-mmWave) MIMO system, which subsumes both IM-assisted and conventional mmWave-MIMO systems as its special cases. More specifically, a set of RF switches are deployed between the ABF and the outputs of RF chains, which determines whether the RF connection works in a conventional or IM-assisted manner. Aided with this generalization, SPIM-mmWave has the potentials to outperform conventional ones.
    
	\item Based on the proposed SPIM-mmWave scheme, we conduct theoretic analysis on the specific design of the RF switches under various channel assumptions, different numbers of antennas, and various signal-to-noise ratio (SNR) regions. Since we mainly focus on the single-user scenario in this paper, we therefore restrict our system setup to a simplified version with only a single RF chain and a single user. Such scenario is simplified but still provides much insight about how IM can help improving the SE.
    
    \item As a conclusion of our analysis, we found that, under a given specific condition with respect to the channel gains and SNR levels, the proposed SPIM-mmWave guarantees to improve the SE of conventional mmWave-MIMO systems. It is worth noting that the corresponding condition is not very stringent in most of the communication scenarios, thus justifying the application of our proposed SPIM-mmWave MIMO systems.
\end{enumerate}

The remainder of this paper is organized as follows. We introduce our proposed SPIM-mmWave system and the corresponding channel models in Section \uppercase\expandafter{\romannumeral2}. In Section \uppercase\expandafter{\romannumeral3} we provide theoretic SE result of the proposed SPIM-mmWave system. In Section \uppercase\expandafter{\romannumeral4}, theoretic analysis is conducted to give the condition under which our proposed system outperforms conventional ones. The simulation results are provided in \uppercase\expandafter{\romannumeral5} to empirically prove our theoretic studies. Finally, Section \uppercase\expandafter{\romannumeral6} concludes this paper.

\textit{Notations:} In this paper, the lowercase and uppercase boldface letters denote column vectors and matrices respectively. The operators $(\cdot)^T$ and $(\cdot)^H$ denote the transposition and conjugate transposition, respectively. $\mathcal{CN}(\bm\mu, \bm\Sigma)$ denotes a circularly symmetric complex-valued multi-variate Gaussian distribution with $\bm\mu$ and $\bm\Sigma$ being its mean and covariance, respectively. $\|\mathbf{M}\|_F$ represents the Frobenius norm of $\mathbf{M}$ and $\vert \mathbf{M} \vert$ is the determinant. $\mathbf{I}_N$ denotes an $N$-dimensional identity matrix. $\text{diag}(\mathbf{v})$ returns a diagonal matrix with diagonal elements given by a column vector $\mathbf{v}$. 

\section{System and Channel Models}
\subsection{Proposed SPIM-mmWave MIMO System Model}
\begin{figure}
	\center{\includegraphics[width=0.90\linewidth]{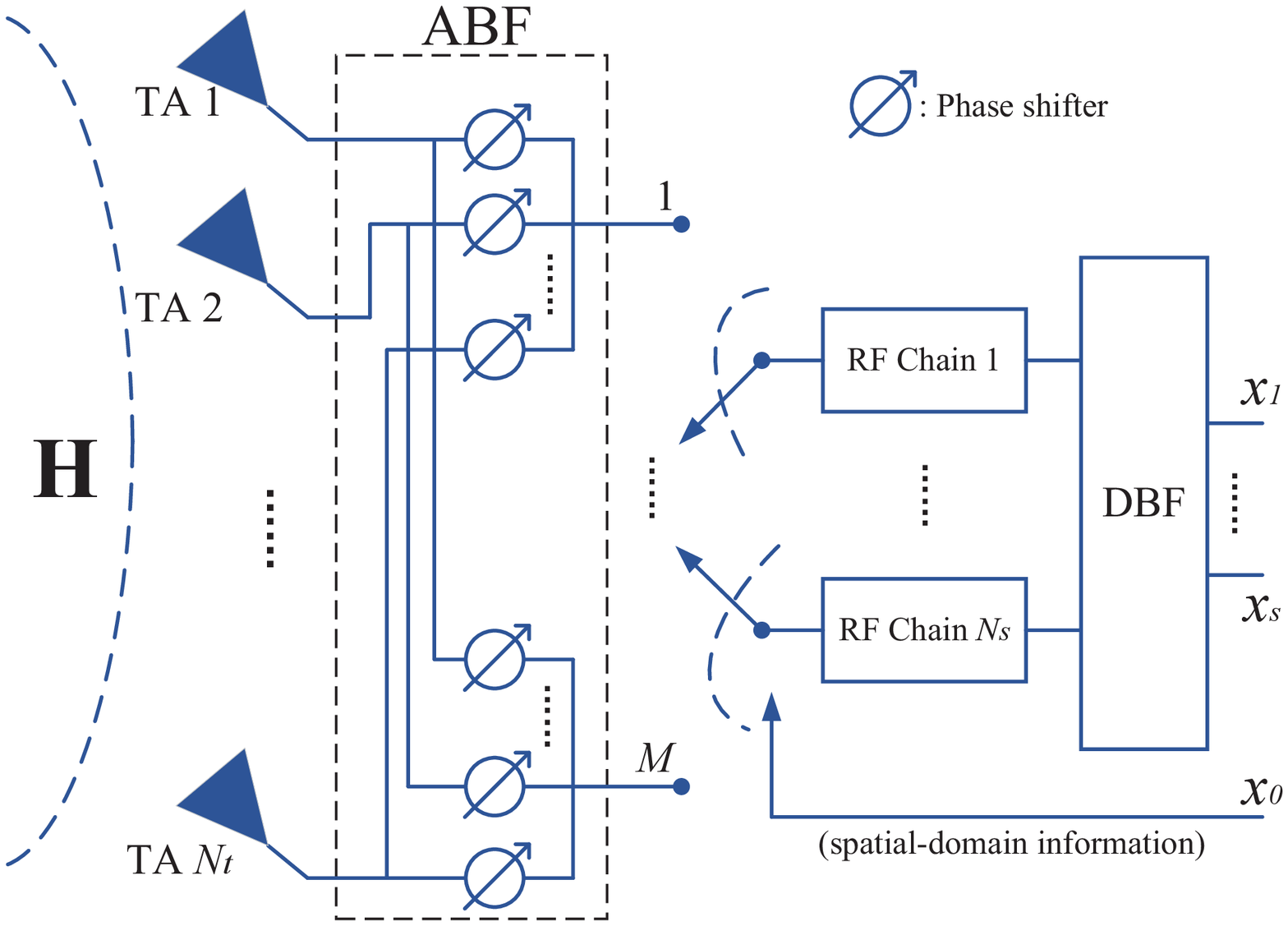}}
	\caption{System model of the proposed SPIM-mmWave MIMO transmitter.}
	\label{Fig_system_model}
\end{figure}

In this paper we consider a point-to-point communication system via mmWave with $N_\text{t}$ transmit antennas (TAs) and $N_\text{r}$ receive antennas (RAs). To better illustrate the proposed system structure, we depict the corresponding system model in Fig. \ref{Fig_system_model}. As can be seen from the figure, the major difference between SPIM-mmWave system and the conventional system is that a set of RF switches are imposed between the outputs of RF chains and the inputs of ABF. For a transmitter with $N_\text{s}\ge 1$ RF chains, we assume that the ABF can maximally process $M\ge N_\text{s}$ inputs, while each of the $M$ inputs is connected to the $N_\text{t}$ TAs via a complete independent set of phase shifters. In this paper, we consider a massive-MIMO scenario, thus implying that $N_\text{t} \gg 1$ as well as $N_\text{t} > M$. For the purpose of convenience, in the remainder of this paper we refer $M$ as the number of \textit{spatial paths}, which represents the maximum number of available spatial transmission channels (but cannot be all \textit{simultaneously} exploited due to the limitation of RF-chain number). We assume that $M \ge N_\text{s}$ throughput this paper. More importantly, in the case of $M > N_\text{s}$, there exists $\binom{M}{N_\text{s}}$ choices of connections between the RF chains and ABF. Therefore we incorporate an extra stream of information, i.e. the \textit{spatial-domain information} (represented by $x_0$, as depicted in Fig. \ref{Fig_system_model}) to randomly assign the outputs of RF chains to the $M$ taps, which is in accordance to the principle of IM, i.e. an information-driven random switching regime. Lastly, we assume that the user takes in $S\ge 1$ independent symbol streams, which are linked to the $N_\text{s}$ RF chain outputs using a $N_\text{s} \times S$ DBF matrix. To fully exploit spatial multiplexing gain, we require that $S=N_\text{s}$ in this paper.

It can thus be seen that, when $M=N_\text{s}$, the proposed SPIM-mmWave MIMO system reduces to a conventional mmWave-MIMO system with HBF as in \cite{gao2016energy} (in this case the spatial-domain information is also absent since there is only one choice of connection). When $M > N_\text{s}$, the proposed system becomes an IM-assisted mmWave-MIMO system as in \cite{he2018spatial}, which literally performs IM with respect to the $M$ spatial paths, in which a part of information is encoded onto the specific indices of spatial paths that are utilized for information transmission. Therefore the proposed system incorporates both conventional and IM-assisted mmWave MIMOs as its special cases.

To facilitate our theoretic analysis, we let $\mathbf{x} \in \mathbb{C}_{N_\text{s}\times 1}$ represent the transmitted symbol vector that is fed to DBF. Similar to many of the current research, it is assumed that $\mathbf{x}$ is subject to a multi-variate complex-valued Gaussian distribution, i.e. $\mathbf{x} \sim \mathcal{CN}(\mathbf{0}, \mathbf{I}_{N_\text{s}})$. Furthermore, we use $\mathbf{D}\in\mathbb{C}_{N_\text{s}\times N_\text{s}}$ and $\mathbf{A}\in\mathbb{C}_{N_\text{t} \times M}$ to represent the DBF and ABF matrices, respectively, thus the signal vector that is transmitted by the $N_\text{t}$ antennas is given by:
\begin{equation}
    \mathbf{s} =\mathbf{A}\mathbf{B}_i \mathbf{D}\mathbf{x}\in\mathbb{C}_{N_\text{t}\times 1},
    \label{signal_expr}
\end{equation}
where $\mathbf{B}_i \in \mathbb{R}_{M \times N_\text{s}}$ represents the $i$-th selected \textit{spatial pattern}, defining the connection between the $N_\text{s}$ RF-chain outputs and the $M$ ABF taps. We assume that $\mathbf{B}_i$ is selected by the spatial-domain information $x_0$ from a spatial patterns' alphabet $\mathcal{B}\triangleq \{\mathbf{B}_1, \ldots, \mathbf{B}_K\}$ with equal probability, while we have 
\begin{equation*}
    K = 2^{\lfloor\log_2 \binom{M}{N_\text{s}}\rfloor},
\end{equation*}
which implies that the spatial domain can at most carry $\lfloor \log_2 \binom{M}{N_\text{s}} \rfloor$ bits of information. Besides, since $\mathbf{B}_i$ defines the spatial connection between RF chains and ABF, we thus require that 
\begin{equation*}
    \mathbf{B}_i = \left[\mathbf{e}_{i_1}, \ldots, \mathbf{e}_{i_{N_\text{s}}}\right],
\end{equation*}
where $\mathbf{e}_{i_t}$ is the $i_t$-th row vector of an identity matrix $\mathbf{I}_{M}$, and $i_t \in [1, M]$ represents the index of ABF taps that the $t$-th RF chain is connected to, and we also require that $i_t \ne i_s$ provided that $t\ne s$. This is because no two RF chains can be connected to the same ABF tap.

\subsection{MmWave-MIMO Channel Model}
Let the $N_\text{r}\times N_\text{t}$ mmWave-MIMO channel model be represented by $\mathbf{H}$. In this paper, we adopt a geometric channel model similar to \cite{ayach2014spatially}-\cite{park2017dynamic}, which is capable of capturing the scattering features of mmWave channels. More specifically, we adopt a narrow-band channel matrix that is formulated as follows,
\begin{equation}
    \mathbf{H} = \mathbf{P} \bm\Lambda \mathbf{Q}^H,
    \label{channel_expr}
\end{equation}
where $\mathbf{P} \in \mathbb{C}_{N_\text{r} \times N_\text{ch}}$, $\bm\Lambda \in \mathbb{C}_{N_\text{ch} \times N_\text{ch}}$, and $\mathbf{Q}\in\mathbb{C}_{N_\text{t} \times N_\text{ch}}$ represent the matrices of receive-antenna response, channel gains, and transmit-antenna response, respectively. Note that we denote the number of effective scattering paths in the mmWave channel as $N_\text{ch}\ge 1$. Moreover, for the $i$-th scattering path ($1 \le i \le N_\text{ch}$), we assume that the distribution of its angle of departure (AoD) and angle of arrival (AoA) are respectively given by  $\hat{\phi}_i \in [-\pi/2, \pi/2]$ and $\hat{\theta}_i \in [-\pi/2, \pi/2]$, then we have
\begin{equation*}
    \arraycolsep=1.0pt\def\arraystretch{1.6}
    \begin{array}{rcl}
        \mathbf{P} &=& \left[\mathbf{a}_\text{R}(\hat{\theta}_1), \ldots, \mathbf{a}_\text{R}(\hat{\theta}_{N_\text{ch}})\right], \\
        \mathbf{Q} &=& \left[\mathbf{a}_\text{T}(\hat{\phi}_1), \ldots, \mathbf{a}_\text{T}(\hat{\phi}_{N_\text{ch}})\right], \\
        \bm\Lambda &=& \text{diag}\left([\sqrt{w_1}, \ldots, \sqrt{w_{N_\text{ch}}}]\right),
    \end{array}
\end{equation*}
where $w_i \ge 0$ ($i\in\{1, \ldots, N_\text{ch}\}$) denotes the $i$-th scattering path's gain. Besides, the transmit and receive antenna response vectors $\mathbf{a}_\text{T}(\hat{\phi}) \in \mathbb{C}_{N_\text{t}\times 1}$ and $\mathbf{a}_\text{R}(\hat{\theta}) \in \mathbb{C}_{N_\text{r}\times 1}$ are respectively given as
\begin{equation*}
	\arraycolsep=1.0pt\def\arraystretch{2.0}
    \begin{array}{rcl}
        \mathbf{a}_\text{T}(\hat{\phi}) &=& \frac{1}{\sqrt{N_\text{t}}}\left[e^{-j\pi\sin\hat{\phi} \left(-\frac{N_\text{t}-1}{2}\right)}, \ldots, e^{-j\pi\sin\hat{\phi} \left(\frac{N_\text{t}-1}{2}\right)}\right]^T, \\
        \mathbf{a}_\text{R}(\hat{\theta}) &=& \frac{1}{\sqrt{N_\text{r}}}\left[e^{-j\pi\sin\hat{\theta} \left(-\frac{N_\text{r}-1}{2}\right)}, \ldots, e^{-j\pi\sin\hat{\theta} \left(\frac{N_\text{r}-1}{2}\right)}\right]^T.
    \end{array}
\end{equation*}

For simplicity, in this paper we use another notations $\phi$ and $\theta$ which are defined as
\begin{equation}
    \phi \triangleq \frac{1}{2}\sin\hat{\phi}, \,\, \theta \triangleq \frac{1}{2}\sin\hat{\theta}.
    \label{phi_theta_expr}
\end{equation}

Therefore, we have $\phi\in[-0.5, 0.5]$ and $\theta \in[-0.5, 0.5]$, as well as
\begin{equation*}
	\arraycolsep=1.0pt\def\arraystretch{2.0}
    \begin{array}{rcl}
        \mathbf{a}_\text{T}(\phi) &=& \frac{1}{\sqrt{N_\text{t}}}\left[e^{-j2\pi\phi \left(-\frac{N_\text{t}-1}{2}\right)}, \ldots, e^{-j2\pi\phi \left(\frac{N_\text{t}-1}{2}\right)}\right]^T, \\
        \mathbf{a}_\text{R}(\theta) &=& \frac{1}{\sqrt{N_\text{r}}}\left[e^{-j2\pi\theta \left(-\frac{N_\text{r}-1}{2}\right)}, \ldots, e^{-j2\pi\theta \left(\frac{N_\text{r}-1}{2}\right)}\right]^T.
    \end{array}
\end{equation*}

Note that, in this paper we assume that $\theta_i$ and $\phi_i$ differs from each other with probability of $1$.

Finally, according to the expressions in (\ref{signal_expr}) and (\ref{channel_expr}), the received signal $\mathbf{y} \in \mathbb{C}_{N_\text{r} \times 1}$ at the receiver is expressed as 
\begin{equation}
    \mathbf{y} = \mathbf{H} \mathbf{A}\mathbf{B}_i \mathbf{D}\mathbf{x} + \mathbf{n},
    \label{transmission_eqn}
\end{equation}
when the $i$-th spatial pattern is chosen. Note that $\mathbf{n} \sim \mathcal{CN}(\mathbf{0}, N_0\mathbf{I})$ represents the additive white Gaussian noise (AWGN).

Similar to conventional mmWave HBF studies, several constraints are imposed on $\mathbf{A}$ and $\mathbf{D}$. First of all, the DBF matrix should be well-conditioned, i.e. $\text{rank}(\mathbf{D}) \ge S$, and is subject to the power constraint
\begin{equation}
    \text{Tr}\left(\mathbf{D}\mathbf{D}^H\right) \le 1.
    \label{constraint_power}
\end{equation}

Second of all, since the ABF is composed of several sets of phase shifters, the components of $\mathbf{A}$ should thus satisfy
\begin{equation}
    \mathbf{A}_{(m, n)} = e^{j\theta_{m,n}}, \,\, 1\le m \le N_t, \,\, 1\le n \le M,
    \label{constraint_phase}
\end{equation}
where the subscript $(m, n)$ denotes the element at the $m$-th row and $n$-th column of a matrix.


\section{Spectral Efficiency Analysis}
\subsection{General Framework}
In order to quantify the system performance of the transmission depicted by (\ref{transmission_eqn}), we exploit the commonly used metric, i.e. mutual information (MI). As a matter of fact, the MI of conventional MIMO systems as well as mmWave-MIMO systems have been well addressed in \cite{goldsmit2003capacity} and \cite{caire2003on}. The theoretic and empirical MI study of IM/SM techniques has also been carried out by various literatures, e.g. \cite{an2015mutual} and \cite{ishikawa2018fifty}. However, the theoretic study of the combination of SM and traditional mmWave-MIMO has always been absent until recent. In \cite{he2018spatial}, the authors proposed an asymptotic SE lower bound for quantifying the SE of SM-assisted mmWave-MIMO systems. The proposed bound can be applied to a variety of such systems with various numbers of antennas, different HBF designs, and different mmWave-MIMO channel realizations. According to \cite{he2018spatial}, the MI of the transmission in (\ref{transmission_eqn}) is given by
\begin{equation}
    I(\mathbf{y}; \mathbf{x}, \mathbf{B}_i) \overset{(a)}{=} I(\mathbf{y}; \mathbf{x}\vert \mathbf{B}_i) + I(\mathbf{y}; \mathbf{B}_i),
    \label{mutual_inf1}
\end{equation}
where $I(\mathbf{y}; \mathbf{x}, \mathbf{B}_i)$ stands for the MI between the received signal $\mathbf{y}$ and the transmitted symbols, i.e. $\mathbf{x}$ and the selected spatial pattern $\mathbf{B}_i$. Note that here $\mathbf{x}$ is a \textit{continuously} random variable (more specifically, a complex-valued multi-variate normal distribution random variable), while $\mathbf{B}_i$ is a \textit{discretely} random variable that is drawn with equal probability from $\mathcal{B}$.

Based on the results of \cite{he2018spatial}, the MI term $I(\mathbf{y}; \mathbf{x} \vert \mathbf{B}_i)$ at the right-hand side of (\ref{mutual_inf1}) represents the conditional MI between $\mathbf{y}$ and $\mathbf{x}$ given that $\mathbf{B}_i$ is acquired, which can be therefore given using the well-known Shannon's formula, i.e.
\begin{equation}
    I(\mathbf{y}; \mathbf{x} \vert \mathbf{B}_i) = \frac{1}{K} \sum_{k=1}^K \log_2\left( \left| \frac{1}{N_0}\bm\Sigma_k \right| \right),
    \label{mutual_inf1_1}
\end{equation}
where $\bm\Sigma_k$ represents the covariance matrix of $\mathbf{y}$ when the $k$-th spatial pattern is selected for transmission. According to (\ref{transmission_eqn}) and the property of multi-variate normal distribution, we have
\begin{equation}
    \bm\Sigma_k = N_0 \mathbf{I} + \mathbf{HAB}_k \mathbf{D}\mathbf{D}^H \mathbf{B}_k^H \mathbf{A}^H \mathbf{H}^H.
    \label{covariance_k}
\end{equation}

Moreover, although the expression of $I(\mathbf{y}; \mathbf{B}_i)$ lacks closed form, the research of \cite{he2017on} has provided us with a closed-form lower bound as follows:
\begin{equation}
    \begin{array}{rcl}
        I_\text{LB}(\mathbf{y}; \mathbf{B}_i) &=& \displaystyle \log_2 K - N_\text{r} \log_2 e - \text{...}\\
        && \displaystyle \frac{1}{K} \sum_{n=1}^K \log_2 \sum_{t=1}^K \frac{\left|\bm\Sigma_n\right|}{\left|\bm\Sigma_n + \bm\Sigma_t\right|},
    \end{array}
    \label{mutual_inf1_2}
\end{equation}
which has been provided as a tight approximation to the real MI term $I(\mathbf{y}; \mathbf{B}_i)$.

Combining (\ref{mutual_inf1}), (\ref{mutual_inf1_1}) and (\ref{mutual_inf1_2}), a closed-form approximation for the MI of (\ref{transmission_eqn}) is thus given by\footnote{Note that a constant gap $N_\text{r}(1-\log_2 e)$ is deleted to neutralize the asymptotic approximation error, as demonstrated in \cite{he2017on}}
\begin{equation}
    \begin{array}{rcl}
        && \displaystyle I_\text{app}(\mathbf{y}; \mathbf{x}, \mathbf{B}_i)  \\
        &=& \displaystyle \log_2 \left(\frac{K}{(2 N_0)^{N_\text{r}}}\right) -  \frac{1}{K} \sum_{n=1}^K \log_2 \sum_{t=1}^K \left|\bm\Sigma_n + \bm\Sigma_t\right|^{-1}.
    \end{array}
    \label{mutual_inf_apprx}
\end{equation}

Note that, due to the limitation of pages, the specific derivations of (\ref{mutual_inf_apprx}) is omitted here and we refer the interested readers to \cite{he2017on} and \cite{he2018spatial} for more technical details.

\subsection{Preliminaries}
\label{Sec_Simplified_Scenario}
\begin{figure*}[t]
    \normalsize
    \begin{equation}
        I_\text{SPIM}(M=2) = \frac{1}{2}\sum_{i=1}^2 \log_2\left(1+\frac{w_i \left\|\mathbf{a}_i\right\|^2}{N_0}\right) + 1 - N_\text{r} - \frac{1}{2} \sum_{n=1}^2 \log_2 \left(\sum_{t=1}^2 \frac{|\bm\Sigma_t|}{|\bm\Sigma_n + \bm\Sigma_t|}\right)
        \label{I_SPIM_expr}
    \end{equation}
    \hrulefill
\end{figure*}

In order to facilitate our theoretic analysis, in this paper we consider a simplified single-user mmWave-MIMO scenario. We summarize our preliminaries as follows.
\begin{enumerate}
    \item We only consider a single RF chain, i.e. $N_\text{s} = S = 1$. In this case, we also have $\mathbf{D} = 1$.

    \item We firstly restrict the maximum number of spatial paths to $N_\text{ch} = M = 2$ and derive a simplified conclusion. Then we will generalize the scattering condition to the case with $N_\text{ch} = M \ge 2$ spatial paths, and derive the condition under which SPIM-mmWave is superior to conventional ones.
    
    \item We assume that the transmitter is equipped with massive antennas, i.e. $N_\text{t} \gg 1$, while the receiver is equipped with multiple RAs ($N_\text{r} > 1$).
\end{enumerate}

\textit{Remark:} It is worth noting that the assumptions of ours is quite practical in common mmWave scenarios. Due to the severe pathloss property of mmWave channels, the number of effective scattering paths could be quite limited. The massive-TA assumption is also practical, considering that the wavelength of mmWave signals is quite small.

\section{Theoretic SE Performance Comparisons}
\subsection{Simplified Case with $M = 2$ Spatial Paths}
\label{subs_simplified}
We commence our analysis from a simplified case of $M=2$, which is easy to derive and could provide us with much insight. In the occasion of $M=2$, we should thus design the transmitter's ABF so that the beams are carefully steered along the $2$ scattering paths with maximum channel gains, i.e.
\begin{equation*}
    \mathbf{A} = \sqrt{N_\text{T}} \left[\mathbf{a}_\text{T}(\phi_1), \,\, \mathbf{a}_\text{T}(\phi_2)\right] \in \mathbb{C}_{N_\text{t}\times 2},
\end{equation*}
which can effectively steer the beams along the directions of $\phi_1$ and $\phi_2$, while fulfilling the requirements of constant amplitude, i.e. (\ref{constraint_phase}).

As we assume that $\phi_1 \ne \phi_2$ with probability of $1$, under the assumption that $N_\text{t} \gg 1$, we thus have
\begin{equation}
    \arraycolsep=1.0pt\def\arraystretch{1.3}
    \begin{array}{rcl}
        \displaystyle \mathbf{HA} & \displaystyle \xrightarrow{N_\text{t} \rightarrow \infty} & \displaystyle \left[\sqrt{w_1 g_1} \mathbf{a}_\text{R}(\theta_1), \sqrt{w_2 g_2} \mathbf{a}_\text{R}(\theta_2)\right] \\
        & = & \displaystyle \left[\sqrt{w_1 g_1} \mathbf{a}_1, \,\, \sqrt{w_2 g_2} \mathbf{a}_2\right],
    \end{array}
    \label{HA_approx}
\end{equation}
where $g_1 = g_2 = N_\text{t}$ denotes the \textit{transmit array gain} provided by ABF.

Besides, given that $M=2$, the corresponding SPIM-mmWave should therefore randomly assign the output of RF chain to one of the $2$ spatial paths, which leads to
\begin{equation}
    \mathbf{B}_i = \mathbf{e}_i, \,\, i \in \{1, \,\, 2\}.
    \label{B_simplification}
\end{equation}

Aided with (\ref{HA_approx}) and (\ref{B_simplification}), the MI approximation in (\ref{mutual_inf_apprx}) can thus be simplified as in (\ref{I_SPIM_expr}), where we also have
\begin{equation*}
    \left|\bm\Sigma_i\right| = N_0^{N_\text{r}} \cdot \left| \mathbf{I}+\frac{w_i g_i}{N_0} \mathbf{a}_i\mathbf{a}_i^H \right| = N_0^{N_\text{r}} \cdot \left(1+\frac{w_i g_i}{N_0}\right),
\end{equation*}
where we have utilized the matrix determinant property, i.e. $|\mathbf{I}+\mathbf{AB}| = |\mathbf{I}+\mathbf{BA}|$, and $\left\|\mathbf{a}_i\right\|^2 = 1$ ($i \in \{1, \,\, 2\}$).

Before proceeding our theoretic analysis, we also need to specify the SE of conventional mmWave system. Under the assumptions we made in Section \ref{Sec_Simplified_Scenario}, it can be easily shown that conventional mmWave system is no more than steering the output of the RF chain to the spatial scattering path with the \textit{maximum channel gain}. Assume that $w_1 \ge w_2$, the received symbol vector in a conventional mmWave-MIMO system is thus expressed as
\begin{equation*}
    \mathbf{y}_\text{mmWave} = x \cdot \sqrt{w_1 g_1} \mathbf{a}_\text{R}(\theta_1) + \mathbf{n} \in \mathbb{C}_{N_\text{r} \times 1},
\end{equation*}
in which we have $x \in \mathcal{CN}(0, 1)$ and $\mathbf{n} \in \mathcal{CN}(\mathbf{0}, N_0\mathbf{I})$. Therefore, according to Shannon's formula, the SE of conventional mmWave systems can thus be given as\footnote{Note that we have achieved \textit{unit transmit power} by applying the aforementioned constraints, thus we have neglected the transmit power and can regard $N_0^{-1}$ as our effective SNR.}
\begin{equation}
    I_\text{mmWave} = \log_2 \left( 1 + \frac{w_1 g_1}{N_0} \right).
    \label{I_mmWave_expr}
\end{equation}

Therefore, in this section we seek to derive the specific channel condition under which $I_\text{SPIM} > I_\text{mmWave}$, where $I_\text{SPIM}$ and $I_\text{mmWave}$ are given by (\ref{I_SPIM_expr}) and (\ref{I_mmWave_expr}), respectively. 

We hereby provide our theoretic result in Theorem \ref{theorem_1}.
\begin{theorem}
    Given that $w_1 \le 4w_2$, the SE achieved by SPIM-mmWave MIMO system guarantees to outperform that of conventional mmWave MIMO systems in the high-SNR region.
    \label{theorem_1}
\end{theorem}

\textit{Proof:} The proof is provided in Appendix \ref{AppA}.

\textit{Remark:} As it can be seen from Theorem \ref{theorem_1}, the theoretic condition under which SPIM-mmWave can outperform conventional mmWave is that the gain of the largest scattering path must not exceed four times that of the second largest path. In some very extreme cases, such as indoor LoS channels, it is possible that $w_1/w_2 \rightarrow \infty$. In such extremely ``imbalanced'' cases, there nearly exists only one spatial path along which the information can be conveyed with acceptable pathloss. Therefore it is of no benefit to exploit SPIM-mmWave techniques. 

However, when the $w_1/w_2$ condition in Theorem \ref{theorem_1} is satisfied, SPIM-mmWave is capable of striking even higher SE performance than its conventional counterpart, of which the reason can be explained as follows. Simply comparing (\ref{I_SPIM_expr}) against (\ref{I_mmWave_expr}), it can be seen that the first term of $I_\text{SPIM}$ is always less than $I_\text{mmWave}$, i.e. 
\begin{equation}
    \frac{1}{2}\sum_{i=1}^2 \log_2\left(1+\frac{w_i}{N_0}\right) \le \log_2 \left(1+\frac{w_1}{N_0}\right),
    \label{I_comparison_1}
\end{equation}
since we have restricted that $w_1 \ge w_2$. From (\ref{I_comparison_1}), it is naturally implied that, via applying SPIM-mmWave, the MI conveyed via the \textit{symbol domain} is always deteriorated, as it devotes half of its transmission period on some weaker scattering path. Nevertheless, such performance loss can be compensated, even over-compensated by the \textit{extra information bits} that IM carries.

\subsection{General Case with $M$ Spatial Paths}
\label{subs_general}
We now consider a more general case with $M$ spatial paths. Similar to Section \ref{subs_simplified}, under the assumption that $N_\text{t} \gg 1$ we have
\begin{equation*}
    \mathbf{HA} \approx \left[\sqrt{w_1 g_1}\mathbf{a}_1, \ldots, \sqrt{w_M g_M}\mathbf{a}_M \right] \in \mathbb{C}_{N_\text{r} \times M}.
\end{equation*}

Besides, for the purpose of brevity, we also assume that $w_1 > w_2 > \ldots > w_M$. In order to utilize the SE approximation in (\ref{mutual_inf_apprx}), we substitute the expression of $|\bm\Sigma_n + \bm\Sigma_t|$ in (\ref{app_1}) into (\ref{mutual_inf_apprx}), which leads to 
\begin{equation*}
    \arraycolsep=1.0pt\def\arraystretch{1.8}
    \begin{array}{l}
        \displaystyle I_\text{SPIM}(M) = \log_2 M - \text{...}\\
        \displaystyle \frac{1}{M}\sum_{n=1}^M\log_2 \sum_{t=1}^M \left[ \left(1+\frac{w_n g_n}{2N_0}\right) \left(1+\frac{w_t g_t}{2N_0}\right) - Q_{n,t}\right]^{-1},
    \end{array}
\end{equation*}
where $Q_{n,t}$ is given as
\begin{equation*}
    Q_{n,t} \triangleq \frac{w_n w_t g_n g_t}{4N_0^2N_\text{r}^2} \cdot \frac{\sin^2\left[ \pi N_\text{r} (\theta_n-\theta_t) \right]}{\sin^2\left[\pi (\theta_n-\theta_t)\right]}.
\end{equation*}

Under the general setup with $M$ spatial paths, our conclusion can be summarized using the following Theorem \ref{theorem_2}.

\begin{theorem}
    For a SPIM-mmWave MIMO system with $M$ spatial paths and a single RF chain to outperform a conventional mmWave-MIMO system in the high-SNR region, the \textit{geometric mean} of the non-LoS components must be greater than $\tau$ times the power of the LoS component, i.e.
    \begin{equation}
        \left(\prod_{n=2}^M w_n\right)^{\frac{1}{M-1}} > \tau \cdot w_1,
        \label{general_condition_1}
    \end{equation}
    where $\tau$ is given by
    \begin{equation*}
        \tau \triangleq M^{\frac{-M}{M-1}} \cdot \exp\left(4N_0 \sum_{n=1}^M w_n^{-1}g_n^{-1}\right).
    \end{equation*}
    \label{theorem_2}
\end{theorem}

\textit{Proof:} The proof is provided in Appendix \ref{AppB}.

If we further increase the SNR, the following corollary can be easily obtained.
\begin{corollary}
    In the region of high SNR, a SPIM-mmWave MIMO system with $M$ spatial paths outperforms a conventional mmWave-MIMO system provided that
    \begin{equation}
        \left(\prod_{n=2}^M w_n\right)^{\frac{1}{M-1}} > M^{\frac{-M}{M-1}} \cdot w_1.
        \label{general_condition_2}
    \end{equation}
    \label{corollary_2}
\end{corollary}

Note that Corollary \ref{corollary_2} can be easily obtained by substituting $N_0$ with $0$ in the conclusion of Theorem \ref{theorem_2}. Besides, if we look at the case where $M=2$, we can easily obtain that $w_2 > w_1 / 4$, which is in exact accordance to Theorem \ref{theorem_1}.

\textit{Remark:} As can be seen from (\ref{general_condition_1}) and (\ref{general_condition_2}), the superiority of SPIM-mmWave against conventional ones depends heavily on the \textit{geometric mean} of the non-LoS components. With no loss of generality, one can assume that the path gains follow the order $w_1 > w_2 > \ldots > w_{N_\text{ch}}$. Therefore, the condition provided by Theorem \ref{theorem_2} and Corollary \ref{corollary_2} requests us to carefully select a ``cut-off'' threshold $M$, so that the paths with very little gains will be not included by SPIM-mmWave and reduce the geometric mean $(\prod_{n=2}^M w_n)^{1/(M-1)}$ significantly. Moreover, we can also look at (\ref{general_condition_2}) via a ``logarithmic'' perspective, i.e.
\begin{equation}
    \frac{1}{M-1}\sum_{n=2}^M \log w_n > C + \log w_1,
    \label{general_condition_log}
\end{equation}
where $C$ is given by $-\frac{M}{M-1}\log M$. Therefore, the condition not only requires that the overall power level of the non-LoS components to be non-negligible, but also prohibits the non-LoS components from having very weak paths, which will result in excessively negative $\log w_n$ and reduce the left-hand side of (\ref{general_condition_log}) significantly.

Finally, we can also assume that the path gains follow an exponentially decaying model, i.e.
\begin{equation}
    w_n = \gamma^{n-1}, n \in \{1, 2, \ldots, M\},
    \label{decaying_paths}
\end{equation}
and assume that there are infinitely many paths we can potentially exploit, i.e. $N_\text{ch} = \infty$ (Although many of the non-LoS paths with very little gains cannot be utilized). 

Besides, we exploit $g_n=N_\text{t}$ as the array gain, and take (\ref{decaying_paths}) into (\ref{general_condition_1}), which leads to
\begin{equation}
    1 < M^{\frac{M}{M-1}} \cdot \gamma^{\frac{M}{2}} \cdot \exp\left[-\frac{4N_0(\gamma^{1-M}-\gamma)}{g_1 (1-\gamma)}\right].
    \label{general_condition_decaying_1}
\end{equation}

With a given decaying exponent $\gamma \in (0, 1)$, we define a novel term \textit{SPIM margin} $M_\text{margin}$ as follows
\begin{equation}
    \arraycolsep=1.0pt\def\arraystretch{1.8}
    \begin{array}{rcl}
        \displaystyle M_\text{margin} &\triangleq& \displaystyle \max_{M>=1} M \\
        &\text{s.t.}& \displaystyle M^{\frac{M}{M-1}} \gamma^{\frac{M}{2}} \cdot \exp\left[-\frac{4N_0(\gamma^{1-M}-\gamma)}{g_1 (1-\gamma)}\right] > 1,\\
        && \displaystyle M = 2^b, \,\, b\in\left\{0, 1, 2, \ldots, b_\text{max}\right\}.
    \end{array}
    \label{M_margin_def}
\end{equation}

Note that here we have restricted $M$ to be $M=2^b$ as a direct requirement of SPIM. Since $M=1$ is always a feasible solution to the optimization in (\ref{M_margin_def}). Therefore, on one hand, if $M_\text{margin}=1$, then conventional mmWave outperforms SPIM-mmWave systems with \textit{arbitrary} value of $M$. On the other hand, if $M_\text{margin}>1$, then a SPIM-mmWave system with $M=2, 4, \ldots, M_\text{margin}$ can all outperform conventional ones. Therefore, the higher $M_\text{margin}$ is, the greater performance improvement can SPIM-mmWave system potentially bring.

\begin{figure}
    \center{\includegraphics[width=0.90\linewidth]{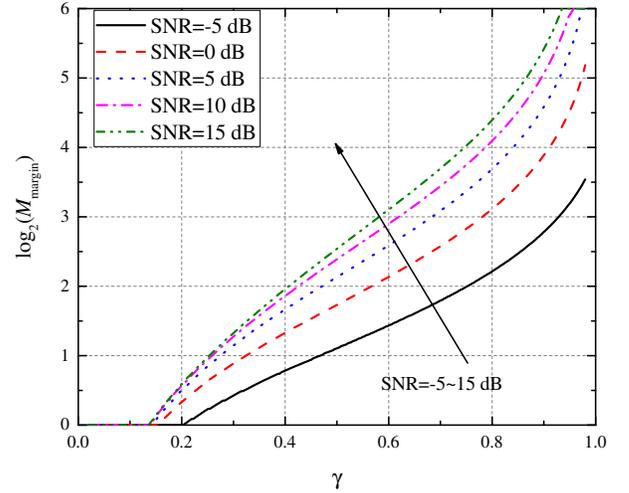}}
    \caption{$\log_2 M_\text{margin}$ as a function of decaying exponent $\gamma$ and various $N_0$ ($1/\text{SNR}$). Besides we have set $b_\text{max}=6$ and relaxed the integer requirement of $\log_2 M$.}
	\label{Fig_Mmargin}
\end{figure}

To provide a more intuitive demonstration on the relation between $\gamma$ and $M_\text{margin}$, in Fig. \ref{Fig_Mmargin} we depict $M_\text{margin}$ as a function of $\gamma$ with various $N_\text{0}$. Note that we have set $b_\text{max}$ to be $6$. Besides, in order to better present the results, we have relaxed the requirement that $M=2^b$ and assume $M$ can be chosen \textit{continuously} from $1$ to $2^{b_\text{max}}$.

As can be seen from Fig. \ref{Fig_Mmargin}, basically $M_\text{margin}$ becomes larger than $1$ when $\gamma$ is higher than $0.2$. More importantly, it is observed that $M_\text{margin}$ increases monotonically with $\gamma$. This is because, as the path gains decay slower, the more available spatial paths there will be. 



\section{Simulation Results}
In this section we provide several numerical simulations to prove the accuracy of our theory. We commence by presenting all our simulation setup in Table \ref{TableSim} for convenience. Hence the simulation parameters of this paper are all configured accordingly unless mentioned otherwise. Note that $U(a, \,\,b)$ in Table \ref{TableSim} represents a uniformly-random distribution over region $(a, \,\,b)$. It is worth noting that, since we consider a single-user system, therefore we have restricted the number of TAs from being too large (e.g. over $100$). However, it can be seen that even with a modest number of TAs, the simulation results are still in great accordance to the theoretic results.

\begin{table}
    \small
    \caption{Simulation Parameters}
    \newcommand{\tabincell}[2]{\begin{tabular}{@{}#1@{}}#2\end{tabular}}
    \centering
    \renewcommand\arraystretch{1.5}
    \begin{tabular}{cll}
    \hline
    Symbols & Specifications & Values \\\hline\hline
    $N_\text{t}$ & No. of transmit antennas & $64$ \\
    $N_\text{r}$ & No. of receive antennas & $8$ \\
    $N_\text{s}$ & No. of RF chains & $1$ \\
    $\phi_i$ & Channel AoDs & $U(-0.35, \,\,0.35)$ \\
    $\theta_i$ & Channel AoAs & $U(-0.25, \,\, 0.25)$ \\
    \hline
    \end{tabular}
\label{TableSim}
\end{table}

\subsection{SE Approximation Accuracy}

\begin{figure}
	\center{\includegraphics[width=0.90\linewidth]{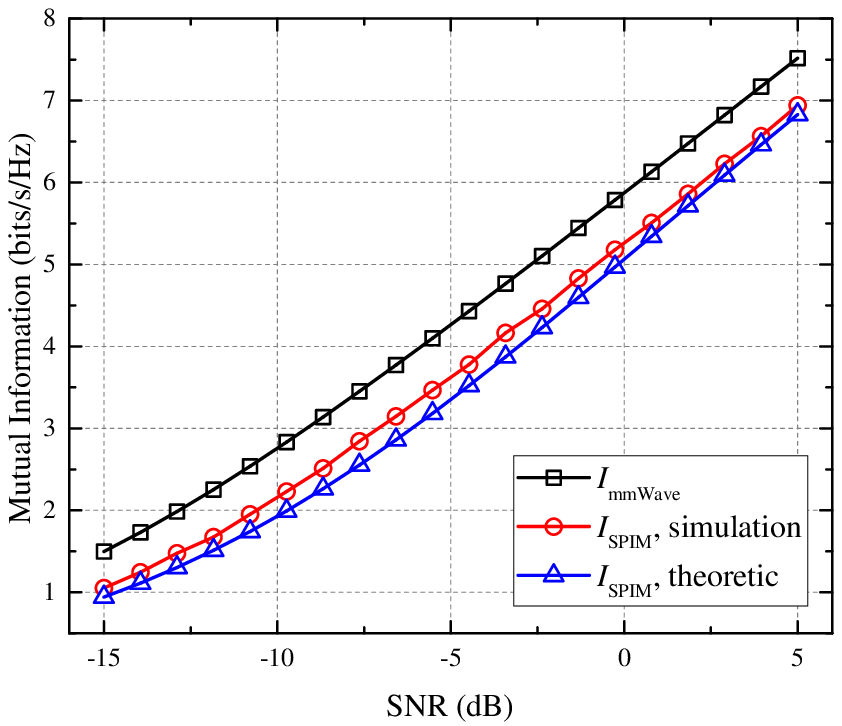}}
	\caption{Mutual information yielded by conventional mmWave and SPIM-mmWave MIMO systems with $(w_1, w_2)=(0.9, 0.1)$.}
	\label{Fig_Graph1}
\end{figure}

\begin{figure}
	\center{\includegraphics[width=0.90\linewidth]{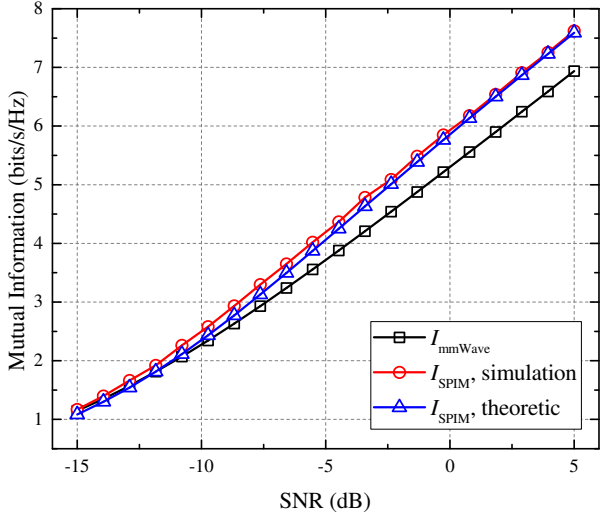}}
	\caption{Mutual information yielded by conventional mmWave and SPIM-mmWave MIMO systems with $(w_1, w_2)=(0.6, 0.4)$.}
	\label{Fig_Graph2}
\end{figure}

We begin by demonstrating the tightness of our SE approximation $I_\text{SPIM}$, i.e. (\ref{I_SPIM_expr}), with respect to the real MI $I(\mathbf{y}; \mathbf{x}, \mathbf{B}_i)$. We denote the real MI term as ``simulation'', while $I_\text{SPIM}$ as ``theoretic'' in the following figures. Note that here $\text{SNR} \triangleq N_0^{-1}$. The simulations are conducted with respect to ``imbalanced'' and ``balanced'' channel conditions, i.e. $(w_1, w_2)=(0.9, 0.1)$ and $(w_1, w_2)=(0.6, 0.4)$, respectively.

As can be seen from Fig. \ref{Fig_Graph1} and Fig. \ref{Fig_Graph2}, the theoretic MI is shown to provide a relatively accurate approximation to the simulation results of MI. More importantly, in the region of high SNR (approximately over $2$ dB), the theoretic MI approximation is almost the same as its true value, which therefore justifies our usage of $I_\text{SPIM}$ in (\ref{I_SPIM_expr}) to quantify the SE performance of SPIM-mmWave system.

Comparing $I_\text{SPIM}$ against $I_\text{mmWave}$, it can be seen that when the channel suffers from an imbalanced condition, e.g. $(w_1, w_2) = (0.9, 0.1)$, the usage of SPIM-mmWave only leads to decreasing the achievable SE. Based on Fig. \ref{Fig_Graph1}, SPIM-mmWave is outperformed by conventional mmWave MIMO by nearly $0.8$ bits/s/Hz. However, when the channel condition is relatively balanced, i.e. $w_1 / w_2 = 3/2$, as in Fig. \ref{Fig_Graph2}, the proposed SPIM-mmWave system is capable of achieving higher SE performance than conventional mmWave MIMOs. In the high-SNR region, an SE improvement of approximately $0.6$ bits/s/Hz is achieved by SPIM-mmWave, according to Fig. \ref{Fig_Graph2}.

\subsection{Performance Guarantee of SPIM-MmWave with $M=2$}

\begin{figure}
	\center{\includegraphics[width=0.90\linewidth]{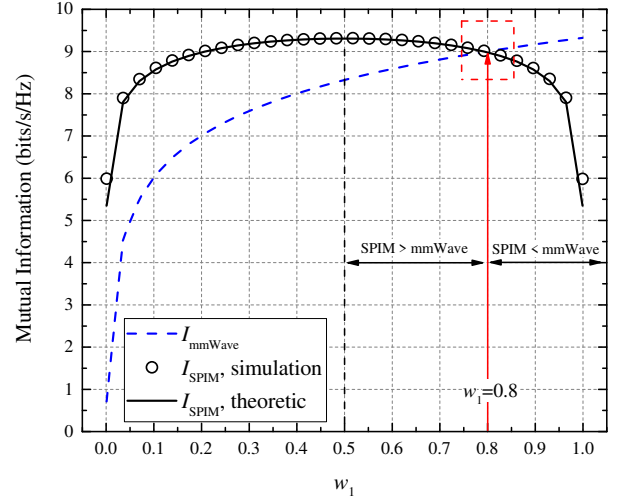}}
	\caption{Mutual information yielded by conventional mmWave and SPIM-mmWave MIMO systems with various $w_1\in [0, 1]$, $w_1+w_2=1$, and $N_0 = 0.1$.}
	\label{Fig_Graph3}
\end{figure}

\begin{figure}
	\center{\includegraphics[width=0.90\linewidth]{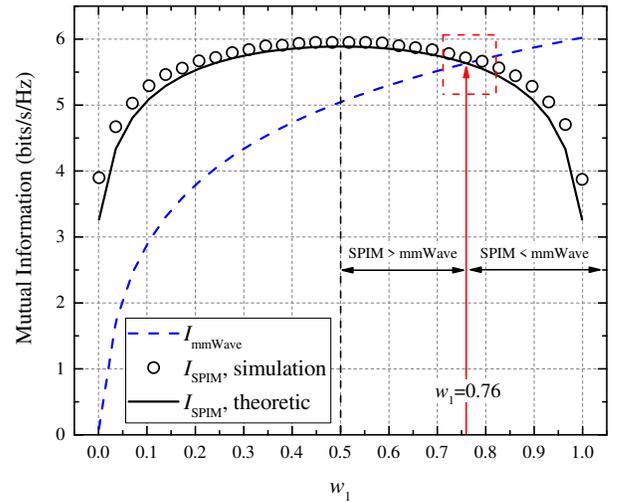}}
	\caption{Mutual information yielded by conventional mmWave and SPIM-mmWave MIMO systems with various $w_1\in [0, 1]$, $w_1+w_2=1$, and $N_0 = 1.0$.}
	\label{Fig_Graph4}
\end{figure}

As can be seen from Theorem \ref{theorem_1}, the performance improvement of SPIM-mmWave relies heavily on the ratio $w_1/w_2$. In order to better illustrate the performance improvement as a function of $w_1/w_2$, we depict the SE yielded by SPIM-mmWave MIMO system (simulation-based MI as well as theoretic approximation $I_\text{SPIM}$) and conventional mmWave MIMO system as a function of $w_1 \in [0, 1]$ in Fig. \ref{Fig_Graph3}, under the constraint that $w_1+w_2=1$.

According to Fig. \ref{Fig_Graph3}, the simulation results and theoretic approximation of $I_\text{SPIM}$ agrees very well to each other. More importantly, it is observed that SPIM-mmWave outperforms conventional ones when $w_1 \le 0.8$ (i.e. $w_1 \le 4 w_2$, since $w_1+w_2=1$), and is outperformed by conventional ones when $w_1 > 0.8$. This is in accordance to our asymptotic analysis of Theorem \ref{theorem_1}, thus proving the accuracy of our theoretic derivations. 

Furthermore, we increase $N_0$ from $0.1$ to $1.0$ and depict the corresponding curves in Fig. \ref{Fig_Graph4}. As can be observed from the figure, with the increase of $N_0$, the ``conversion point'' of $w_1$ becomes lower than $0.8$. As a matter of fact, SPIM-mmWave maintains its performance superiority only when $w_1 \le 0.76$, i.e. the performance advantage of SPIM-mmWave can be deteriorated by the increasing of noise. However the overall conclusion is not changed qualitatively, that is, SPIM-mmWave is preferred when the non-LoS channel components are non-negligible.

\subsection{Performance of SPIM-MmWave with $M$}

\begin{figure}
    \center{\includegraphics[width=0.90\linewidth]{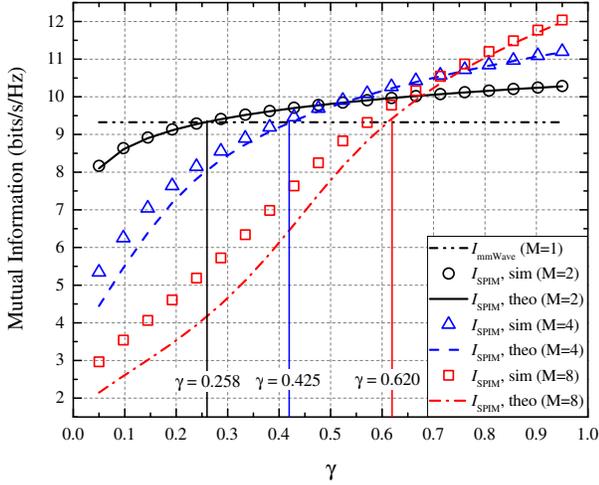}}
    \caption{Mutual information yielded by conventional mmWave and SPIM-mmWave MIMO systems with various $\gamma \in (0, 1)$ and $M\in\{1, 2, 4, 8\}$. $N_0 = 0.1$.}
	\label{Fig_Graph_decay}
\end{figure}

Finally, we look into the performance of conventional mmWave and SPIM-mmWave MIMO systems with various $\gamma$ and $M$, where $\gamma$ represents an exponential decaying factor among the path gains, as defined in (\ref{decaying_paths}). According to Fig. \ref{Fig_Graph_decay}, under the influence of $\gamma$, the channel gains decay more slowly with the increase of $\gamma$ from $0$ to $1$. With the increase of $\gamma$, the application of SPIM-mmWave MIMO system with higher value of $M$ is preferred with respect to maximizing the achievable SE. For example, when $\gamma>0.75$, SPIM-mmWave with $M=8$ is the optimal selection, while $M=1$ (i.e. conventional mmWave) is the optimal selection only when $\gamma<0.25$. 

Besides, it is also observed that the simulation and theoretic results of $I_\text{SPIM}$ match each other relatively close, although some deviations still present when $M$ is large. More importantly, we have also marked the three $\gamma$ values that is calculated based on (\ref{general_condition_decaying_1}), i.e. the minimum $\gamma$ value above which a SPIM-mmWave system with $M$ spatial paths outperforms conventional mmWave, which should satisfy the following equation,
\begin{equation*}
    1=M^{\frac{M}{M-1}} \cdot \gamma^{\frac{M}{2}} \cdot \exp\left[-\frac{4N_0(\gamma^{1-M}-\gamma)}{g_1(1-\gamma)}\right],
\end{equation*}
and we denote the $\gamma$ satisfying the above equation by $\gamma(M)$ for convenience. Although $\gamma(M)$ lacks a closed-form solution, we can still numerically estimate it via Newton's method. The calculation results are $\gamma(2) \approx 0.258$, $\gamma(4) \approx 0.425$ and $\gamma(8) \approx 0.620$. As can be seen by Fig. \ref{Fig_Graph_decay}, the estimated $\gamma(M)$ agrees very well with the simulation results, which therefore proves the accuracy of our analysis.

\section{Conclusions}
In this paper, we proposed a novel SPIM-mmWave MIMO scheme as a combination of mmWave and IM. Aided with the information-driven random-switching principle of IM, the proposed SPIM-mmWave MIMO system constitutes a generalized version of mmWave MIMO, and therefore has the potential to outperform conventional ones. Furthermore, we have also conducted thorough analysis to provide a theoretic channel condition under which the proposed SPIM-mmWave MIMO system outperforms conventional ones. Based on our analysis, it is observed that SPIM-mmWave is preferred when the channel path gains are not too imbalanced, e.g. in a LoS scenario. To sum up, SPIM-mmWave MIMO provides a novel perspective to compensate, even outperform the current mmWave-MIMO systems in many common channel conditions.

\appendices
\section{Proof of Theorem \ref{theorem_1}}\label{AppA}
\begin{IEEEproof}
    We begin by re-formulating the term $|\bm\Sigma_t|/|\bm\Sigma_n + \bm\Sigma_t|$ in (\ref{I_SPIM_expr}). Suppose that $n \ne t$, we thus have
    \begin{equation*}
        \arraycolsep=1.0pt\def\arraystretch{1.6}
        \begin{array}{rcl}
            && \displaystyle \left|\bm\Sigma_n + \bm\Sigma_t\right| \\
            &=& \displaystyle \left| 2N_0 \mathbf{I} + w_n g_n \mathbf{a}_n \mathbf{a}_n^H + w_t g_t \mathbf{a}_t\mathbf{a}_t^H \right| \\
            &=& \displaystyle (2N_0)^{N_\text{r}} \cdot \left|\mathbf{I} + \frac{1}{2N_0} \left(w_n g_n\mathbf{a}_n\mathbf{a}_n^H + w_t g_t\mathbf{a}_t\mathbf{a}_t^H\right)\right| \\
            &=& \displaystyle (2N_0)^{N_\text{r}} \cdot \left| \mathbf{I} + \mathbf{R}^H \mathbf{R}\right|,
        \end{array}
    \end{equation*}
    where $\mathbf{R}$ is given by
    \begin{equation*}
        \mathbf{R} \triangleq \left[\sqrt{\frac{w_n g_n}{2N_0}}\mathbf{a}_n, \,\, \sqrt{\frac{w_t g_t}{2N_0}}\mathbf{a}_t\right] \in \mathbb{C}_{N_\text{r}\times 2}.
    \end{equation*}

    Therefore, we have
    \begin{equation*}
        \arraycolsep=1.0pt\def\arraystretch{2.2}
        \begin{array}{rcl}
        && \displaystyle \left|\bm\Sigma_n + \bm\Sigma_t\right| = (2N_0)^{N_\text{r}} \cdot \text{...}\\
        && \displaystyle \left[\left(1+\frac{w_n g_n}{2N_0}\right) \left(1+\frac{w_t g_t}{2N_0}\right) - \frac{1}{4N_0^2} \left|\mathbf{h}_n^H \mathbf{h}_t\right|^2\right],
        \end{array}
    \end{equation*}
    where $\mathbf{h}_n \triangleq \sqrt{w_n g_n} \mathbf{a}_\text{R}(\theta_n)$. Substituting $|\bm\Sigma_n+\bm\Sigma_t|$ into $|\bm\Sigma_t|/|\bm\Sigma_n+\bm\Sigma_t|$, we have
    \begin{equation}
        \arraycolsep=1.0pt\def\arraystretch{1.6}
        \begin{array}{rcl}
            && \displaystyle \frac{|\bm\Sigma_t|}{|\bm\Sigma_n + \bm\Sigma_t|} \\
            &=& \displaystyle \frac{2^{-N_\text{r}} \left(1+ \frac{w_t g_t}{N_0}\right) }{\left(1+\frac{w_n g_n}{2N_0}\right) \left(1+\frac{w_t g_t}{2N_0}\right)-\frac{1}{4N_0^2} |\mathbf{h}_n^H \mathbf{h}_t|^2}.
        \end{array}
        \label{app_1}
    \end{equation}

    Moreover, the inner product $\mathbf{h}_n^H \mathbf{h}_t$ can also be further elaborated as
    \begin{equation*}
        \arraycolsep=1.0pt\def\arraystretch{2.2}
        \begin{array}{rcl}
            \mathbf{h}_n^H \mathbf{h}_t &=& \displaystyle \frac{\sqrt{w_n w_t g_n g_t}}{N_\text{r}}\sum_{i=0}^{N_\text{r}-1}e^{-j 2\pi (\theta_t-\theta_n)[i-(N_\text{r}-1)/2]} \\
            &=& \displaystyle \frac{\sqrt{w_n w_t g_n g_t}}{N_\text{r}} e^{j\pi (N_\text{r}-1)(\theta_t-\theta_n)} \sum_{i=0}^{N_\text{r}-1} e^{-j2\pi i(\theta_t-\theta_n)} \\
            &=& \displaystyle \frac{\sqrt{w_n w_t g_n g_t}}{N_\text{r}} \cdot \frac{\sin\left[\pi N_\text{r} (\theta_t-\theta_n)\right]}{\sin\left[\pi (\theta_t-\theta_n)\right]}.
        \end{array}
    \end{equation*}

    Replacing $\mathbf{h}_n^H \mathbf{h}_t$ into (\ref{app_1}), we thus have
    \begin{equation}
        \arraycolsep=1.0pt\def\arraystretch{1.6}
        \begin{array}{rcl}
            && \displaystyle \frac{|\bm\Sigma_t|}{|\bm\Sigma_n + \bm\Sigma_t|} \\
            &=& \displaystyle \frac{2^{-N_\text{r}} \left(1+ \frac{w_t g_t}{N_0}\right)}{\left(1+\frac{w_n g_n}{2N_0}\right) \left(1+\frac{w_t g_t}{2N_0}\right) - \frac{w_n w_t g_n g_t}{4N_0^2 N_\text{r}^2} \cdot \frac{\sin^2\left[\pi N_\text{r} (\theta_t-\theta_n)\right]}{\sin^2\left[\pi (\theta_t-\theta_n)\right]}}.
        \end{array}
        \label{app_3}
    \end{equation}

    Let $I_\text{SPIM} = I_\text{mmWave}$, we thus have
    \begin{equation}
        \frac{N_0 + w_1 g_1}{4\left(N_0 + w_2 g_2\right)} = \exp\left(-\frac{2}{1+\gamma w_1w_2g_1g_2}\right),
        \label{app_4}
    \end{equation}
    where we have
    \begin{equation*}
        \arraycolsep=1.0pt\def\arraystretch{2.8}
        \begin{array}{rcl}
            \gamma &\triangleq& \displaystyle \frac{1-g_\theta/N_\text{r}^2}{2N_0^2} \left(2+\frac{1}{N_0}\right)^{-1},\\
            g_\theta &\triangleq& \displaystyle \frac{\sin^2\left[\pi N_\text{r} (\theta_t-\theta_n)\right]}{\sin^2\left[\pi (\theta_t-\theta_n)\right]}
        \end{array}
    \end{equation*}

    Finally, considering the asymptotically high-SNR case, i.e. $N_0 \rightarrow 0$, we thus have $\gamma \rightarrow \infty$, which leads to 
    \begin{equation*}
        \frac{N_0 + w_1 g_1}{4\left(N_0 + w_2 g_2\right)} = 1 \Rightarrow w_1 = \frac{3N_0}{N_\text{t}} + 4w_2 \approx 4 w_2,
    \end{equation*}
    where we have also applied $g_1 = N_\text{t}$. Therefore, our conclusion in Theorem \ref{theorem_1} is proved.
\end{IEEEproof}

\section{Proof of Theorem \ref{theorem_2}}\label{AppB}
\begin{IEEEproof}
    We commence by providing an intuitive demonstration of the property of function $Q(\Delta \theta)$ with various $N_\text{r}$ in Fig. \ref{Fig_Q_func}, which is defined as
    \begin{equation*}
        Q(\Delta \theta) = \frac{\sin^2(\pi N_\text{r}\Delta \theta )}{N_\text{r}^2 \sin^2(\pi \Delta \theta)}.
    \end{equation*}

    \begin{figure}
        \center{\includegraphics[width=0.90\linewidth]{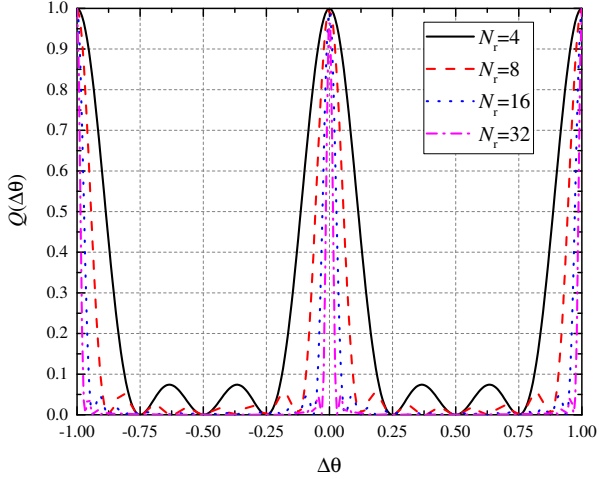}}
        \caption{$Q(\Delta \theta)$ as a function of various $\Delta \theta$ and $N_\text{r}$.}
        \label{Fig_Q_func}
    \end{figure}

    As can be seen from Fig. \ref{Fig_Q_func}, when $N_\text{r}$ is large, $Q_{n,t}$ thus becomes significant only when $\theta_n-\theta_t$ is around $0$, $1$ or $-1$. Taking (\ref{phi_theta_expr}) into consideration, the cases $Q_{n,t}>0$ actually corresponds the cases where $\hat{\theta}_n=\hat{\theta}_t$, $\hat{\theta}_n=-\hat{\theta}_t=\pi/2$ or $\hat{\theta}_n=-\hat{\theta}_t=-\pi/2$. Therefore, in order to simplify our theoretic analysis, we assume that the difference of angles, i.e. $\theta_n-\theta_t$, lies in the region where $Q_{n,t}$ is small enough to be neglected. Note that such assumption becomes asymptotically solid with the increase of $N_\text{r}$. 
    
    Therefore we have
    \begin{equation*}
        \arraycolsep=1.0pt\def\arraystretch{2.8}
        \begin{array}{rcl}
            && \displaystyle I_\text{SPIM}(M) = \log_2 M - \frac{1}{M}\sum_{n=1}^M \log_2 \left[ \left(1 + \frac{w_n g_n}{N_0}\right)^{-1} + \text{...}\right.\\
            && \displaystyle \left.\sum_{t \ne n} \left(1+\frac{w_n g_n}{2N_0}\right)^{-1} \left(1+\frac{w_t g_t}{2N_0}\right)^{-1} \right],
        \end{array}
    \end{equation*}

    Moreover, we have
    \begin{equation*}
        \arraycolsep=1.0pt\def\arraystretch{2.4}
        \begin{array}{rcl}
            \displaystyle I_\text{SPIM}(M) &=& \displaystyle \log_2 M - \frac{1}{M} \sum_{n=1}^M \log_2\left\{\left(1+\frac{w_ng_n}{N_0}\right)^{-1}  \cdot \text{...} \right.\\
            &&\displaystyle  \left. \left[1 + \sum_{t \ne n} \frac{2N_0+2w_ng_n}{2N_0+w_ng_n} \left(1+\frac{w_t g_t}{2N_0}\right)^{-1}\right] \right\} \\
            &\overset{(a)}{\approx}& \displaystyle \log_2 M + \frac{1}{M}\sum_{n=1}^M \log_2 \left(1+\frac{w_n g_n}{N_0}\right) - \text{...}\\
            && \displaystyle \frac{1}{M}\sum_{n=1}^M \log_2\left[1+2\cdot \sum_{t\ne n}\left(1+\frac{w_t g_t}{2N_0}\right)^{-1}\right],
        \end{array}
    \end{equation*}
    where (a) holds when we only consider the high-SNR region, i.e. $N_0 \ll 1$. 
    
    Continue applying the high-SNR assumption and apply the following approximation
    \begin{equation*}
        \log_2(1+x) \approx x\cdot \log_2 e, \,\, x \rightarrow 0,
    \end{equation*}
    which further simplifies $I_\text{SPIM}$ to be
    \begin{equation*}
        \arraycolsep=1.0pt\def\arraystretch{2.4}
        \begin{array}{rcl}
            I_\text{SPIM} &\approx& \displaystyle \log_2 M + \frac{1}{M}\sum_{n=1}^M \log_2 \left(1+\frac{w_n g_n}{N_0}\right) - \text{...}\\
            && \displaystyle \frac{2\log_2 e}{M} \sum_{n=1}^M\sum_{t\ne n}\left(1+\frac{w_t g_t}{2N_0}\right)^{-1} \\
            &\overset{(a)}{\approx}& \displaystyle \log_2 M + \frac{1}{M}\sum_{n=1}^M \log_2\left(1+\frac{w_n g_n}{N_0}\right)-\text{...} \\
            && \displaystyle \frac{4N_0(M-1)\log_2 e}{M} \sum_{t=1}^M w_t^{-1}g_t^{-1},
        \end{array}
    \end{equation*}
    where (a) is again yielded by applying the high-SNR approximation. Let $I_\text{SPIM} \le I_\text{mmWave}$ in which $I_\text{mmWave}$ is given in (\ref{I_mmWave_expr}), the following inequalities can be obtained.
    \begin{equation}
        \arraycolsep=1.0pt\def\arraystretch{2.4}
        \begin{array}{rcl}
            && \displaystyle \log_2 M - \rho\sum_{n=1}^M g_n^{-1}w_n^{-1} \\
            &\le& \displaystyle -\frac{1}{M}\sum_{n=2}^M \log_2\left(1+\frac{w_n g_n}{N_0}\right)+\frac{M-1}{M}\log_2\left(1+\frac{w_1 g_1}{N_0}\right),
        \end{array}
        \label{mutual_inf_ineq}
    \end{equation}
    where $\rho$ is given by
    \begin{equation*}
        \rho = \frac{4N_0(M-1)\log_2 e}{M}.
    \end{equation*}

    After several manipulations of (\ref{mutual_inf_ineq}), we have
    \begin{equation*}
        \sum_{n=2}^M\log_2\left(\frac{N_0+w_1 g_1}{N_0 + w_n g_n}\right) \ge M\log_2 M - M \rho \sum_{n=1}^M w_n^{-1} g_n^{-1}.
    \end{equation*}

    Finally, we apply the high-SNR approximation, which leads to
    \begin{equation*}
        \arraycolsep=1.0pt\def\arraystretch{2.4}
        \begin{array}{rcl}
        w_1 &\ge& \displaystyle \left(\prod_{n=2}^M w_n\right)^{\frac{1}{M-1}} \cdot \text{...}\\
        && \displaystyle  M^{\frac{M}{M-1}} \cdot \exp\left( -4N_0\sum_{n=1}^M w_n^{-1}g_n^{-1} \right).
        \end{array}
    \end{equation*}

    In other words, for SPIM-mmWave system to outperform conventional mmWave-MIMO systems, the \textit{geometric mean} of the non-LoS components, i.e. $(\prod_{n\ne 1} w_n)^{1/(M-1)}$, must be greater than $\tau w_1$ with $\tau$ given by
    \begin{equation*}
        \tau \triangleq M^{\frac{-M}{M-1}} \cdot \exp\left(4N_0 \sum_{n=1}^M w_n^{-1}g_n^{-1}\right),
    \end{equation*}
    and therefore our proof is concluded.
\end{IEEEproof}

\end{document}